\documentclass[10pt,letterpaper]{article}

\usepackage{cogsci}

\cogscifinalcopy 

\usepackage{booktabs}
\usepackage{graphicx} 
\usepackage{amsmath}
\usepackage{pslatex}
\usepackage{apacite}
\usepackage[authoryear]{natbib}
\usepackage{booktabs} 
\usepackage{float} 

\usepackage{xcolor}
\usepackage{hyperref}
\usepackage{listings}
\hypersetup{
    colorlinks,
    linkcolor={red!50!black},
    citecolor={blue!50!black},
    urlcolor={blue!80!black}
}
\usepackage{amsfonts}
\newcommand\blfootnote[1]{%
  \begingroup
  \renewcommand\thefootnote{}\footnote{#1}%
  \addtocounter{footnote}{-1}%
  \endgroup
}
\global\hyphenpenalty=1000
\usepackage{makecell} 

\usepackage{listings}
\lstdefinestyle{prompt}{
  basicstyle=\ttfamily\small,
  breaklines=true,
  columns=fullflexible,
  keepspaces=true,
  frame=single
}

\usepackage[most]{tcolorbox} 

\newtcblisting{promptbox}{
  listing only,
  colback=white,
  boxrule=0.6pt,
  left=2mm,right=2mm,top=1mm,bottom=1mm,
  listing options={
    basicstyle=\ttfamily\small,
    breaklines=true,
    columns=fullflexible,
    keepspaces=true
  }
}

\renewcommand{\thefootnote}{$\dagger$}

\title{Generation and Evaluation in the Human Invention Process \\ through the Lens of Game Design}

 \author{ \\ 
 {
 \large\bf Katherine M. Collins*\textsuperscript{1,2}}, \quad 
 {\large\bf Graham Todd*\textsuperscript{3}}, \quad 
 {\large\bf Cedegao E. Zhang\textsuperscript{2}}, \quad 
{\large\bf Adrian Weller\textsuperscript{1}}, \\ 
{\large\bf Julian Togelius\textsuperscript{3}},  \quad 
{\large\bf Junyi Chu\textsuperscript{4}}, \quad 
{\large\bf Lionel Wong\textsuperscript{2,4}},
{\large\bf Thomas L. Griffiths$\dagger$\textsuperscript{5}},\quad 
{\large\bf Joshua B. Tenenbaum$\dagger$\textsuperscript{2}} \\ \\
\textsuperscript{1}University of Cambridge \quad
\textsuperscript{2}MIT\quad
\textsuperscript{3}NYU \quad \textsuperscript{4}Stanford University \quad \textsuperscript{5}Princeton University \\
Correspondence to \texttt{katiemc@mit.edu}, \, \texttt{gdrtodd@nyu.edu}
 }

\begin{document}

\maketitle

\begin{abstract}

The human ability to learn rules and solve problems has been a central concern of cognitive science research since the field’s earliest days. But we do not just follow rules and solve problems given to us by others: we modify those rules, create new problems, and set new goals and tasks for ourselves and others. Arguably, even more than rule following and problem solving, human intelligence is about creatively breaking and stretching the rules, changing the game, and inventing new problems worth thinking about. Creating a good rule or a good problem depends not just on the ideas one can think up but on how one evaluates such proposals. Here, we study invention through the lens of game design. We focus particularly on the early stages of novice, ``everyday'' game creation, where the stakes are low. We draw on a dataset of over $450$ human created games, created by participants who saw an initial seed set of two-player grid-based strategy games. We consider two different cognitive mechanisms that may be at work during the early processes of intuitive game invention: an associative proposal based on previous games one has seen and compute-bounded model-based evaluation that an everyday game creator may use to refine their initial draft proposals. In our preliminary work, we conduct a model-based analysis of how people invented new games based on prior experience and find that generated games are best described by a model which incorporates model-based estimates of game quality at a population level. Our work points to how human invention is based not only on what people propose, but how they evaluate and offers a computational toolkit to scale empirical studies of model-based simulation in open-ended human innovation. 

\end{abstract}

\section{Introduction}

Much of individual and collective behavior is governed by systems of rules and reward, from governments to sports leagues and from markets to family policies. But people do not merely shape their actions to accommodate existing systems and incentive structures -- people regularly decide to \textit{change} the rules and create entirely new systems in the process.
People change and make new rules in a variety of contexts, whether defining new kickoff rules in American football or amending a constitution to enshrine a new legal right. These modifications can also be composed with each other and copied into other systems, a process that enables the evolution of ancient games into their modern forms or the spread of effective legislation around the globe. People's tendency to modify systems of rules extends into their personal lives~\citep{daston2023rules}, and even young children are sensitive to changes in rule systems and reward structures \citep{rakoczy2008taking, rakoczy2008sources,  schulz2012finding, chu2020play, rule2023fun}.

\blfootnote{* indicates equal first author contribution. $\dagger$ indicates equal senior author contribution. Presented at CogSci 2025.}

What accounts for people's ability to effectively generate novel systems of rules and flexibly navigate the resulting unfamiliar problems? In this paper, we study the invention of new systems of rules through the lens of \textit{game design}.
Games are instructive in part because they provide a microcosm of complex social behavior \citep{Suits1978-SUITGG} while often remaining simple to describe and modify.
Game play also has a long history in cognitive science and AI as a model of planning, decision making under uncertainty, and theory of mind~\citep{  cleveland1907psychology, newell1958chess, crowley1993flexible, gobet2004moves, lake2017building, van2023expertise, allen2023using, chase1973mind}.
Here, we focus on the ways in which people alter existing games or concoct new games altogether to better suit their whims or preferences. What kinds of cognitive mechanisms are recruited in this generative process?

\begin{figure*}[h!]
    \centering
    \includegraphics[width=1.0\linewidth]{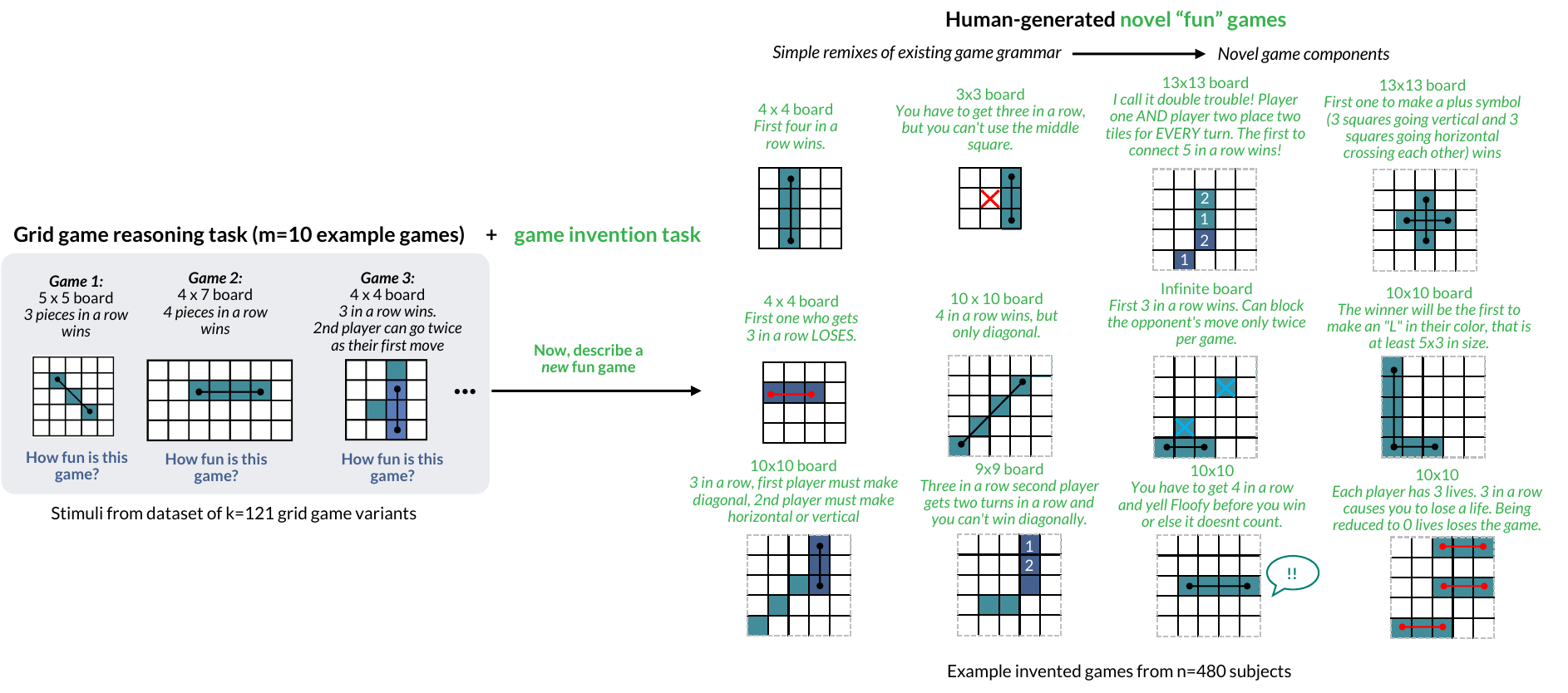}
    \vspace{-0.8cm}
    \caption{\textbf{Overview of the invented game dataset collection process.} Participants first rated a series of two-player grid based strategy games on either how fun or fair they would be (see \cite{collins2026intuitivegamer}). Participants then created a new game they thought would be fun to play. People created a wide range of new games (examples are presented on the right).} 
    \label{fig:example-games}
\end{figure*}

Our main hypothesis is that even novice game designers go beyond either random generation or merely recombining previously seen instances when proposing new games. Rather, we posit that ``everyday'' innovation involves both a \textit{proposal} process that synthesizes previously-encountered examples and an \textit{evaluative} process that actually simulates the dynamics and likely outcomes of the generated game or system. The role of compositionality and context in human invention is well-studied \citep{youn2015invention, davidson2022creativity, zhao2024rational}, but there are a number of challenges that stand in the way of computationally assessing the role of mental simulation on people's process of invention: (1) quantifiably evaluating a new system requires the ability to translate from its description (presumably in natural language) to a more structured or programmatic scheme on which simulations can be performed; and (2) an ``evaluative'' designer would need to be able to reason about the likely moves of potential players despite, by definition, never having played the game themselves.


Here, we present a new set of techniques for addressing the challenges outlined above.
First, we tackle the problem of representation by leveraging the ability of modern language models to translate between natural language and structured world models in code \citep{wong2023word}.
In our case, we use these models to convert natural language game descriptions into the recent \texttt{Ludax} domain-specific language that represents a wide class of board games as simulable environments \citep{todd2025ludax}.
To address the challenge of simulating behavior, we take advantage of a recent empirically verified computational model of human play in novel games \citep{collins2026intuitivegamer}. This \textit{Intuitive Gamer} model uses simple and game-general goal-directed heuristics to probabilistically choose reasonable moves on even completely unfamiliar games; this model is a better predictor of novice player actions than both  computationally more sophisticated and more naive alternatives. Taken together, these techniques allow us to implement a cognitively-plausible evaluative process with which we can retroactively score human-generated games at scale. 

We apply this collection of modeling techniques to understanding people's creative processes using a new dataset of over 450 human-created board games. Our descriptive results indicate that novice designers \textit{are} affected by the kinds of games they have recently been exposed to, often sampling and recombining previously-seen game mechanics rather than inventing them whole-cloth. Context is not everything, however: several participants created games with entirely novel mechanics not seen in any previous examples.
Our model-based results indicate that a sampling process that incorporates the evaluation scores, derived from simulating under the Intuitive Gamer model, is a better fit to the human data than one which accounts only for a game's likelihood in the context of other recently-seen exemplars.
In addition, we find that the games participants generated (whether recombining existing mechanics or inventing new ones) tend to be scored as more ``fun'' under the Intuitive Gamer model than the original $121$ games the games that participants would have seen as context or games otherwise sampled from the space of possible games. Taken together, our results offer preliminary evidence for the role of model-based simulation in everyday innovation.




Our proposal connects to and expands on a broader body of work in cognitive science and AI that combines initial proposals with structured evaluations, to model creativity and invention with varying amounts of mental resources ~\citep{bonawitz2010deconfounding, bourgin2014empirical, abbott2015random, ullman2016coalescing, suchow2017evolution, lieder2020resource, davidson2024goals, campbell1960blind}, as well as work on the automatic generation of novel games \citep{pell1992metagame, togelius2008experiment, browne2010evolutionary, cook2011multi, khalifa2017general, todd2024gavel}. The combination of Intuitive Gamer modeling over symbolic representations in \texttt{Ludax} -- which can be concretized from open-ended natural language -- opens up new possibilities into the study of ``everyday'' innovation and paves the way for work that more directly interrogates the cognitive mechanisms at work in the human invention process.

\section{Invented games dataset}

We consider the dataset of $N=484$ novel games collected by ~\cite{collins2026intuitivegamer}. Each game, described in natural language, was contributed by a unique participant who had seen a series of other natural language games from the same class of two-player grid-based strategy games (see Figure~\ref{fig:example-games}). Specifically, participants, recruited from Prolific~\citep{palan2018prolific}, first judged a series of $11$ variants of grid-based games: Tic-Tac-Toe and $10$ other games of varying board sizes and rules sampled from a larger bank of $120$ examples (e.g., ``\textit{second player can play twice},'' ``\textit{first person to $3$ in a row loses}''). Participants were asked to rate each of these games without playing them on either ``funness'' and fairness. They were then instructed to create a new game that they would find fun to play. Participants were required to spend at least one minute before submitting their game and were provided with an optional interactive scratchpad of a possible game board. After at least one minute had passed, participants were instructed to type in the board size and win conditions for their proposed game in two separate text boxes on the same page.
There was no upper limit on the amount of time participants could spend creating their game, and participant compensation was not related to the resulting game's quality.
After creating their game, participants rated the game on how fun or fair they thought the new game they proposed would be (based on whether they had previously rated funness or fairness for the initial set of $11$ games).

\paragraph{Coding invented games} 
To better understand the ways in which the games participants created are affected by those they saw, we manually code the participant-created games according to the high-level semantic categories used by \cite{collins2026intuitivegamer}. Specifically, we encode whether each game contained \textit{constrained win} conditions (e.g., only horizontal or only diagonal lines), \textit{asymmetric win} conditions between the first and second player, \textit{inverted win} conditions (or misère-style games), or the ability for players to make \textit{multiple moves} in a row. In addition, we coded whether the game was a simple \textit{M-N-K} game and whether it involved making a specific shape (e.g. ``\textit{make 2 $\times$ 2 squares}'') and we note that each game could occupy multiple categories. 

\section{Game invention computational model}

\begin{figure*}[t!]
    \centering
    \includegraphics[width=1.0\linewidth]{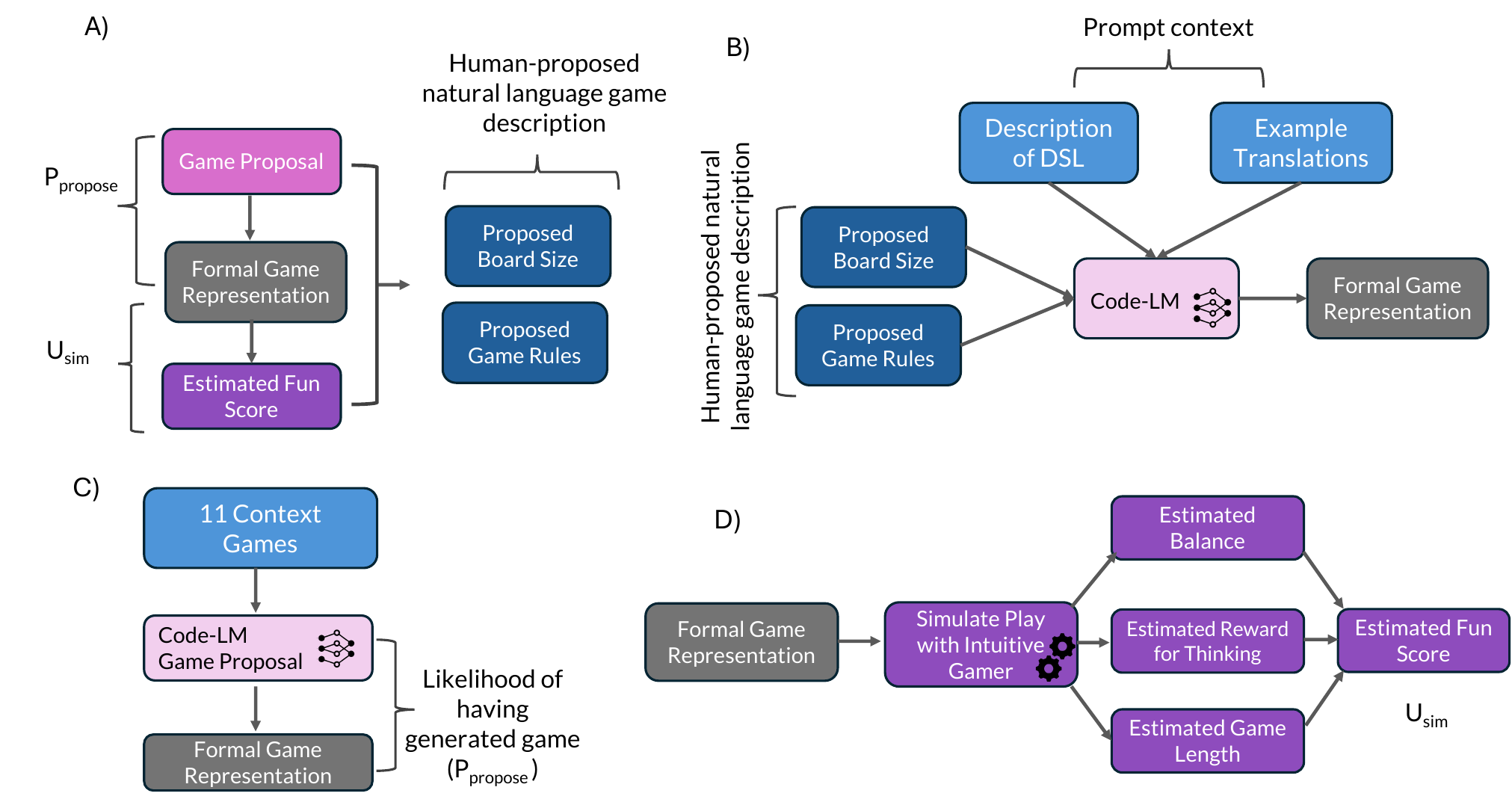}
    \caption{\textbf{Language-computable open-ended game invention model.} (A) Proposed two-stage pipeline for human game invention, where games are created with the goal of creating a fun game. We posit people first sample a game from a proposal distribution over the space of games ($P_\text{propose}$) and then evaluate the game to estimate its likely funness ($U_\text{sim}$) before returning their proposed game. 
    Game details (board size and rules) are expressed in natural language, however, to support simulation, we assume these require a formal representation intermediary. (B) We only have direct access to people's provided games in natural language (i.e., the board size and the rules), which we ``formalize'' into a structured symbolic form (represented in \texttt{Ludax}). Games are formalized using language-to-code models (a ``Code-LM''), where the LM is prompted with a description of \texttt{Ludax} and several example translations. This step allows our pipeline to be ``language-computable'' (i.e., can operate over freeform natural language games). (C) To estimate $P_\text{propose}$ for any one human-created game, we score the formalized representation of the human-created game under a (separate) Code-LM conditioned on the games that people had seen. The context games are presented to the Code-LM in their formalized \texttt{Ludax} form (see Supplement). (D) We use the Intuitive Gamer model~\citep{collins2026intuitivegamer} to simulate how people may reason about the game in their first encounter over the \texttt{Ludax} representation. From those simulated games, a series of general game reasoning features are computed over the simulated play (an estimate of how balanced the game is; how rewarding the game is to think about; and how long the game may take to play). These readouts are combined to give a single to produce an ``estimated funness'' score ($U_\text{sim}$).} 
    \label{fig:method-overview}
\end{figure*}

How did people come up with novel and potentially engaging games after being exposed to a set of examples?  We posit that people engage two mechanisms: (1) an informal, context-conditioned proposal stage in which participants generate descriptions of games in natural language that informally ``sound like'' the ones that were seen, and (2) a more structured formalization and evaluation stage in which participants actually simulate the mechanics and dynamics of the proposed game (Figure~\ref{fig:method-overview}A). This ``propose and refine via evaluation'' model is similar to ones that have been described in earlier work as noted above, which frame novel brainstorming as recombination over a grammar~\citep{ullman2016coalescing}, but relaxes the proposal stage to explain a key feature of real human game generation: people do not only propose new games that are ``mere'' recombinations of the formal semantics of games they've already seen. By modeling the proposal stage as a context-conditioned proposal distribution in natural language, we broaden this stage to draw on prior knowledge about what games ``tend to sound like'' while also conditioning closely on the specific games that people have seen in recent context.

And while people describe their games in natural language, thinking might be cast as unfolding over a formal symbolic representation~\citep{fodor1975the, wong2023word, davidson2022creativity}. Representing games with a formal representation is critical to support simulated play in the proposed games (Figure~\ref{fig:method-overview}A). Here, we only have direct access to elicited natural language game description from people and therefore need to infer a reasonable corresponding formal representation (Figure~\ref{fig:method-overview}B). We next describe these three stages: proposal, evaluation, and formalization in more detail.

\paragraph{Game proposal ($P_{\text{proposal}}$) and formalization}

People often create something new, in part, by drawing on what they have seen before~\citep{campbell1960blind, youn2015invention, davidson2022creativity, zhao2024rational}. However, a full model of human invention needs to be able to not only draw on what was seen, but go \textit{beyond} prior experience (Figure~\ref{fig:method-overview}C). Take, for instance, ``shape-completion'' games like \textit{``The first person to make an L shape consisting of 5 pieces wins. The L shape must be either vertical or horizontal, diagonal Ls do not count.}'' This kind of game (variants of which appear multiple times throughout our dataset) seem very plausible under a general prior over board game rules but would not be covered by the space of games participants were exposed to.

Language models (LMs) offer a reasonable tool to approximate a general proposal distribution that can both be arbitrarily conditioned on past examples (e.g., the games participants saw) and draw on other background knowledge to act as a reasonable prior over whether a game may be ``fun.'' Moreover, distributional LMs trained jointly on language \textit{and code} (``Code-LMs'') offer a method to bridge the freeform games that participants generated with symbolic representations over which simulations can be performed. A core part of our hypothesis is that people also evaluate the games (or more broadly, new systems and problems) they create as part of inventing them. Here, we hypothesize that such evaluation involves simulated play. In order to simulate play across the arbitrary space of games that people propose, we need some way of representing the game formally. Hence, we incorporate a \textit{formalization} step which to interchange between natural language and formal symbolic representations of games that defines the game's environment, transition dynamics, and win conditions. (Figure~\ref{fig:method-overview}B-C). 

Ideally, this representation will also be efficient to operate over; we use the \texttt{Ludax} domain-specific language (DSL)~\citep{todd2025ludax}, which is itself inspired by the general \texttt{Ludii} game description language~\citep{piette2020ludii}. Games in \texttt{Ludax} are defined by formal specifications of \textit{equipment} (i.e. the shape and size of board used), \textit{play rules} (specifying how players take moves), and \textit{end rules} (specifying how the game's winner or loser is determined). \texttt{Ludax} does not encode \textit{all} possible games people have ever created, nor all human-generated games in our dataset (see Supplement). Some games, such as those that are played in real time instead of by taking turns, cannot be presented in \texttt{Ludax}. We present an example of the DSL code for a participant-generated game in \autoref{fig:dsl-example} and further examples in the Supplement. With this formalized symbolic representation, we can then score human-created games under a Code-LM to assess how likely they may be under a proposal distribution conditioned on the games people had seen (Figure~\ref{fig:method-overview}C).


\begin{figure}[t!]
    \centering
    \includegraphics[width=\linewidth]{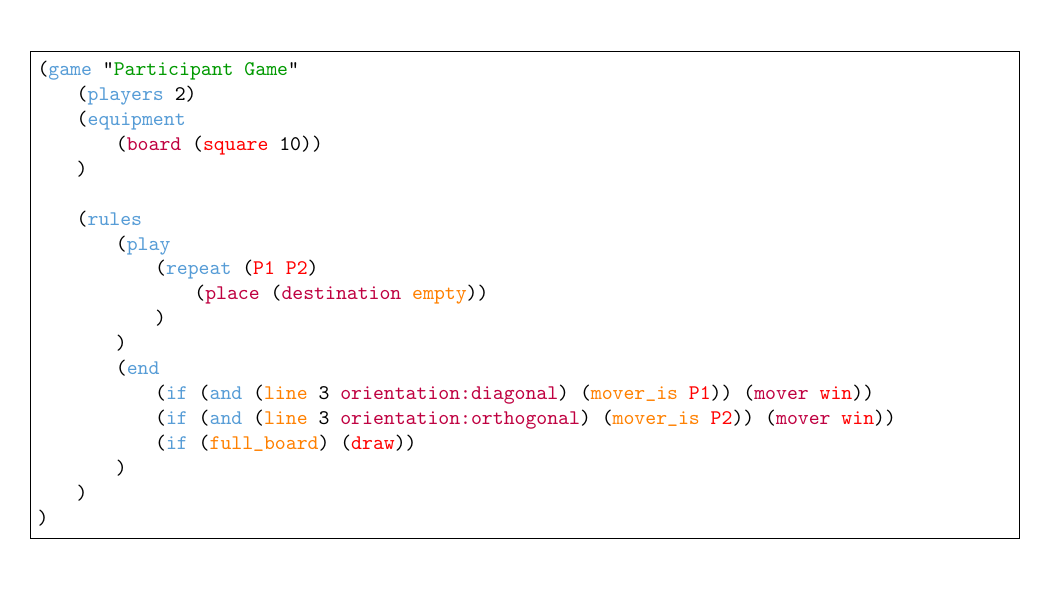}
    \caption{\textbf{The rules for a participant-generated game, encoded into our DSL.} The code here roughly translates as follows: ``the game takes place on a $10 \times 10$ board. Player 1 and Player 2 alternate placing a piece into an empty square. If the first player makes a line of three in a diagonal direction, they win. If the second player makes a line of three vertically or horizontally, they win. If there are no empty spaces, the game is a draw.''}  
    \label{fig:dsl-example}
\end{figure}

\paragraph{Game evaluation using Intuitive Gamer model simulations  ($U_\text{sim}$)}

The second phase of our hypothesis is that people engage in evaluation -- specifically model-based simulation -- when assessing whether their invention meets the goal (i.e., here, to make a fun game). Under the assumption that game inventors would not have played their proposed games (as part of creating a fundamentally ``new'' game), the model used to simulate play ought to capture \textit{novice} player behavior. 

To that end, we consider a structured evaluation of a game using simulations under a ``fast and flat, goal-directed probabilistic'' Intuitive Gamer model of human play and novice game reasoning behavior~\citep{collins2026intuitivegamer}. To obtain an estimate of a game's quality, we apply their same Intuitive Gamer model to simulate what compute-bounded novice reasoning about a new game may be like, based on a simulated play. Here, we make a few important generalizations to be able to cover a broader range of the kinds of novel games people create. \cite{collins2026intuitivegamer} consider only games that are won or lost by completing a single line of particular length and orientation. Under this class, goal progress can be computed as the largest $n$-in-a-row obtained by the action that could still be extended to finish the game (i.e., not already blocked by the other player or the edge of the board). We expand the modeled class of games to not only include arbitrary pattern completion (e.g., ``\textit{You win if you make a plus sign consisting of 5 squares}'') but also the logical composition of multiple conditions (e.g., ``\textit{You lose if you make 3 in a row or if you make a 2-by-2 square of pieces''}). 

Using this model, we can obtain an estimate of a game's funness by computing a series of readouts to estimate features of the games that we posit relate to the game's funness: whether the game is balanced, rewarding to think about, and not too long or too short, following ~\citep{collins2026intuitivegamer}. These readouts can then be combined to estimate a single ``funness'' score ($U_{\text{sim}}$) for the game (Figure~\ref{fig:method-overview}D). We plan to explore other game evaluations that do not include model-based simulation, e.g., just assessing whether games ``sound like'' fun from the natural language game description alone in future work (see ``Alternate explanations and limitations'' section).




\paragraph{Model-based likelihood analysis}

Together, the proposal distribution over games coupled with an evaluation procedure define a generative model over games. In this work-in-progress, we take preliminary steps to explore the role of different cognitive mechanisms that people may deploy in generating novel games, specifically a proposal distribution and evaluative scoring function as described above. However, a challenge in performing this analysis is that our data is ``presence only.'' That is, we only have access to the small number of samples that participants submitted and not to any of the potentially larger set of games they considered but discarded. This methodological challenge is common in ecological modeling, where researchers have access only to limited space of observations sampled from a much larger geographical area \citep{phillips2006maximum}. A common technique for overcoming this problem is Maximum Entropy (MaxEnt) modeling, a tool for assessing the relative contribution of different features in explaining observed data relative to a broad space of possible samples. The technique has its roots in information theory \citep{jaynes1957information} and has also been used in cognitive modeling to assess how people navigate the space of possible programs \citep{ho2018human}. In our case, samples are games ($g$) expressed in the \texttt{Ludax} description language and the larger observation space is the set of all possible games ($\mathcal{G}$).

We define a scoring function that expresses the ``full'' generative model over games, by taking into account both the probability under the proposal distribution ($P_\text{proposal}$) and the Intuitive Gamer-based simulated funness score ($U_{\text{sim}}$), weighted by some parameter $\theta$:

\[
f(g_i; \theta) = \log P_{\text{base}}(g_i) + \theta \cdot U_{\text{sim}}(g_i).
\]

\noindent The MaxEnt model then specifies the probability: 

\[
P(g_i \mid \theta) = \frac{\exp\left(f(g_i; \theta)\right)}{Z(\theta)} = \frac{\exp\left(\log P_{\text{base}}(g_i) + \theta \cdot U_{\text{sim}}(g_i)\right)}{Z(\theta)},
\]

\noindent where $Z$ is the normalization over the sampled space of possible games $\mathcal{G}$: 
\[
Z(\theta) = \sum_{g' \in \mathcal{G}} \exp\left(f(g'; \theta)\right).
\]

\noindent The full space of $\mathcal{G}$ intractably large, and we must also necessarily restrict our analysis to the subset of $\mathcal{G}$ expressible in \texttt{Ludax}. To obtain a somewhat representative collection of such games to use in computing the normalization factor $Z$, we use a heuristic-based sampling strategy described below (see ``Implementation Details'').

The above gives a probability of any given produced game $g_i$. We next define the distribution over the dataset of $N$ human-created games ($D$) (assuming games are sampled i.i.d.) as follows: 

\[
\log \mathcal{L}(\theta; \mathcal{D}) = \sum_{i=1}^N f(g_i; \theta) - N \cdot \log Z(\theta).
\]

\noindent Following \cite{phillips2006maximum}, we incorporate L1 regularization over $\theta$, therefore maximizing the following likelihood objective over the human-created games:

\[
\log \mathcal{L}(\theta; \mathcal{D}) = \sum_{i=1}^N f(x_i; \theta) - N \cdot \log Z(\theta) - \lambda \cdot |\theta|.
\]

\begin{figure*}[t!]
    \centering
        \includegraphics[width=0.95\linewidth]{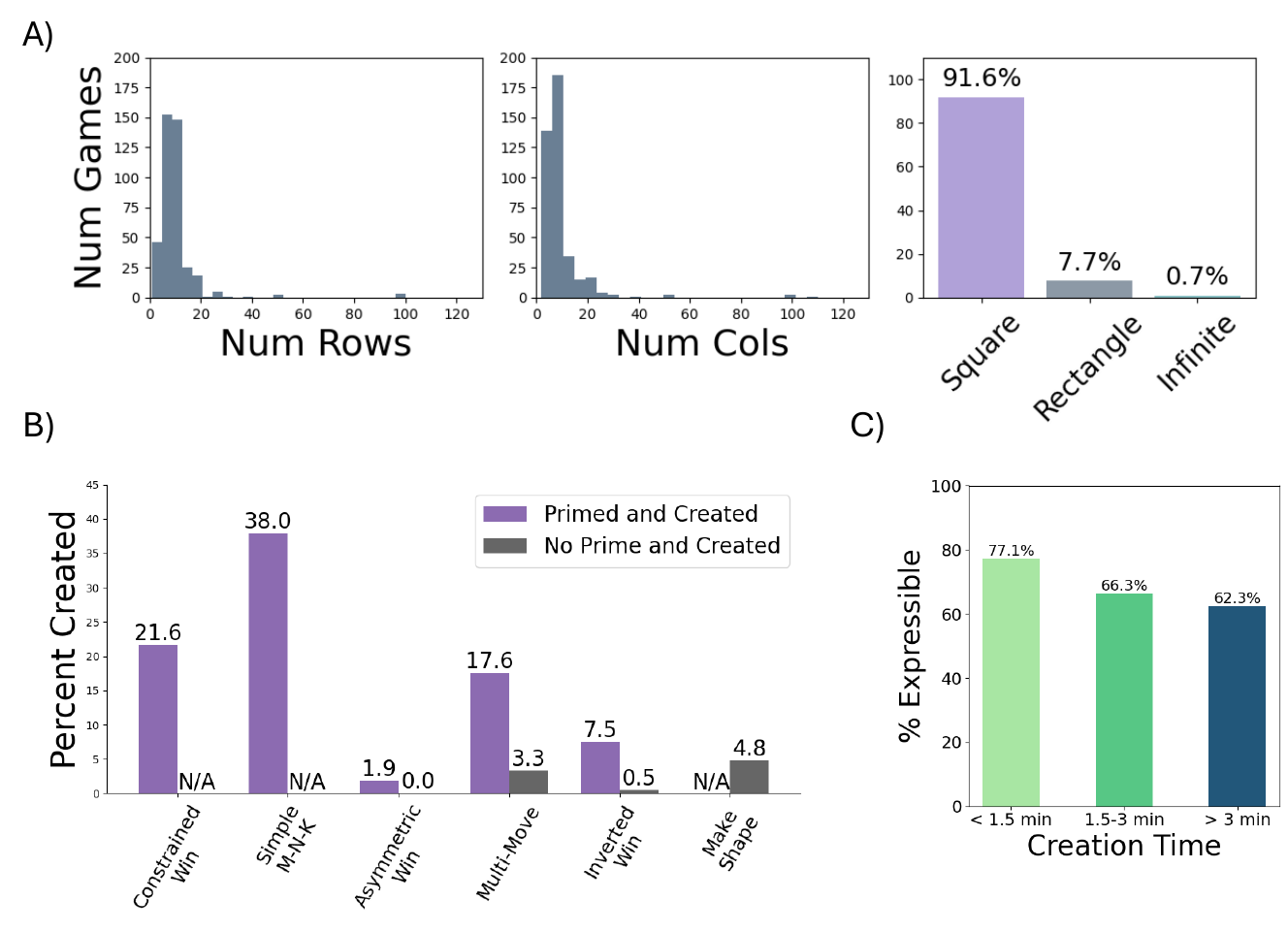}
    \vspace{-0.3cm}
    \caption{\textbf{Characterizing participant-created games.} (A) Board size and shape; (B) Coded game rules and dynamics, as they relate to the kinds of games people saw before creating their game (everyone judged $11$ other games first). Purple indicates the \% of people who created a game with a feature they had previously seen; grey indicates the \% of people who created a game with a feature that did \textit{not} appear in their $11$ examples. N/A indicates that either no one was \textit{not} prompted with an example of that game (i.e., people always saw at least one ``simple'' M-N-K game and one Constrained Win variant) or were never prompted with a game of that type (e.g., no one saw any game that had ``shapes'' other than straight lines as part of win conditions). (C) Percent of human-created games that are expressible in the restricted grammar that is used to represent the original $121$ games, broken up by the amount of time that a participant took to create their game ($N=153$, $193$, and $138$ per bucket, respectively).}
    \label{fig:game-shape}
\end{figure*}

\paragraph{Implementation details}


$P_\text{proposal}$ is computed as the average token log probabilities of the concretized \texttt{Ludax} program representation, conditioned on (for people) the games that each participant had seen, or for the other games, a randomly-sampled set of a similarly $11$ game context (from the same base of $121$ games). Token probabilities are computed under LLaMA 3.1 8B~\citep{dubey2024llama}. 

To compute $U_\text{sim}$, we use simulated self-play playouts under the Intuitive Gamer model \citep{collins2026intuitivegamer} in order to assess the role of model-based evaluation in game generation. We do so only over games that can be formalized into \texttt{Ludax}. We run simulations over $402$ ($\sim 83\%$) of the participant-generated games converted to \texttt{Ludax} programs via our formalization process described above (games were manually excluded if they could not possibly be expressed in \texttt{Ludax}; we also left out $45$ games with boards larger than $12 \times 12$ to avoid very long evaluation times). Games are formalized using LLaMA 3.3 70B (see Supplement). 

In order to obtain the normalization factor $Z$ defined above, we sample $1000$ possible games in $\mathcal{G}$ (which we refer to as ``Random''; see Supplement for details on the sampling process). Games are sampled directly in the formal symbolic \texttt{Ludax} space. We include the original original $121$ games and the subset of people's games that are expressible in \texttt{Ludax} as part of the normalization to cover an approximate space of games. We also simulate game reasoning under the Intuitive Gamer model over all $121$ games as well as the $1000$ randomly-sampled games, in order to compute the normalization factor. 





\section{Results}

We next assess the kinds of games people created -- how those games are influenced by context and how they may be explained by simulating play when assessing potential funness -- using a mix of descriptive and model-based analyses under an initial version of our proposed generative model over games.




\paragraph{Most people created moderately-sized, square boards.}

We first inspect the bare ``environment'' that people proposed: the shape of the board. We find evidence that people are not searching uniformly through the space of recombinations over a game grammar. The vast majority of the games people propose are moderately-sized, square boards (the median proposed board size is $9 \times 9$ and approximately 92\% are square, see Figure ~\ref{fig:game-shape}A). This choice may reflect patterns related to both components of our models -- square boards may be favored in the prior (e.g., common games like Go, Tic-Tac-Toe, or Chess are played on square boards) and small boards may be more likely to lead to games that are not too long (favored in the Intuitive Gamer funness model). On the other hand, these results also demonstrate that -- even when it comes to this basic feature -- people are not simply recombining the attributes of games that they already saw. The average median board size participants saw was approximately $5.5 \times 6.6$ (excluding ``infinite'' boards) and roughly $15\%$ of the games featured non-square boards. Approximately 11\% of participants created a game with a board larger than any game they had seen in the study. 


\paragraph{Many people created games that combined what they had seen, yet some innovators proposed entirely new rule types.}

We next explore the relationship between the specific kind of rules participants saw and what they created. In general, participants were more likely to create games with atypical game mechanics (i.e., mechanics not found in Tic-Tac-Toe) if they were exposed to games with the same coded ``mechanic type'' in the study (\autoref{fig:game-shape}B). Approximately $18\%$ ($53$ of $301$) of participants who saw games involving one player moving two pieces on their turn created a game that also had a multi-move structure, compared to only approximately $3\%$ ($2$ of $59$) of participants who did not see that game type. Similarly $7\%$ ($13$ of $173$) created a game with a win inversion if they saw such a game (e.g., \textit{``4 $\times$ 4. 3 pieces in a row loses''}) compared to only $1$ of $188$ who did not. 

However, some participants also created truly novel games that were not expressible as recombinations or modifications of any of the games they encountered in the study: several participants ($4.8\%$) created games based around forming a specific shape (e.g. an ``L'' or plus-sign) despite not having been primed in the $11$ examples with any win condition other than making a single connected contiguous line. We continue this preliminary analysis of novelty by using a restricted version of the \texttt{Ludax} grammar. Using the LM-based formalization procedure described above, we convert each game to a \texttt{Ludax} program (Figure~\ref{fig:method-overview}B).
We then parse the programs according to a version of the grammar designed to span only the space covered by the original set of $121$ games used by \cite{collins2026intuitivegamer} (what we call the ``restricted grammar'' -- Supplement Figure~\ref{fig:method-overview}) to give a more formal sense of whether the invented game can be expressed as a modification of the games seen during the study. In an exploratory analysis, we find that game creators who took longer tended to create more novel games relative to the set of games they saw, as determined by whether the translation is representable in the restricted grammar (Figure ~\ref{fig:game-shape}C). These results further lend support to the role of context in innovation and the capacity for people to go beyond mere recombination~\citep{kirton1976adaptors}. 






\begin{figure*}[t!]
    \centering

    \includegraphics[width=1.0\linewidth]{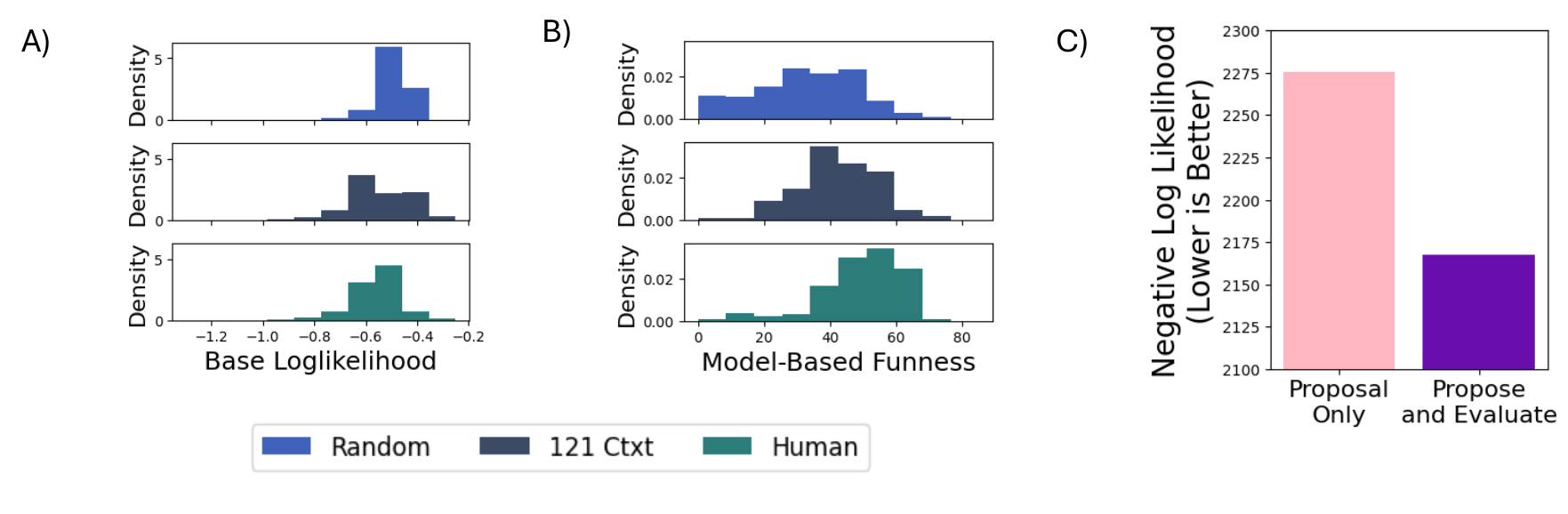}

    \vspace{-0.5cm}
    \caption{\textbf{Likelihood of human-generated games under game creation processes.} (A) Likelihood of each game's \texttt{Ludax} program representation under the base proposal (average token log probability, conditioned on the original batches seen) for games sampled randomly from a DSL over games (DSL; top), the original batch of $121$ games (middle), and the new games created by people (bottom). The log probability is computed over each game, conditioned on example games that either the participant saw or were randomly sampled from the set of $121$ (for top and middle). Log probability is computed using the prompt participants had seen of the experiment instructions and creation task (see Supplement). (B) Predicted funness under our Intuitive Gamer model-based simulator ($U_\text{sim}$). (C) Aggregate negative log likelihood of human-generated games under a model that only accounts for $P_\text{propose}$ (pink) versus the aggregate likelihood under a two-stage model that incorporates $U_\text{sim}$ (purple). Lower is better, highlighting the generally better fit of the two-stage model toward explaining the human-generated games.} 
    \label{fig:likelihood-score}
\end{figure*}

\paragraph{Participants' created games are best explained by a model which accounts for simulated gameplay.}


Using the formalized representation of games, we next assess the likelihood of having generated the human-created games under our hypothesized two stage pipeline. We compare a likelihood model that just incorporates the score of games under the proposal ($P_\text{propose}$) compared to a model that accounts for evaluation that a novice game designer may deploy to assess whether their game might actually be fun ($U_\text{sim}$). Human-created games are comparably likely under $P_\text{propose}$ to alternate games that may be sampled from the space of possible games (Figure~\ref{fig:likelihood-score}A). What distinguishes the games people proposed? The games people proposed are generally predicted to be more fun compared to the context games or games that may otherwise have been sampled randomly from the space of possible games (Figure~\ref{fig:likelihood-score}B). 

To begin to assess the potential contribution of model-based simulation in explaining the games people create, we score the likelihood of the games people created under the different modeling components ($P_\text{propose}$ only; or $P_\text{propose}$ with simulation-based evaluation of gameplay under $U_\text{sim}$) in a MaxEnt model as outlined above. The latter model (incorporating both stages of our proposed pipeline) is parameterized by a single weight, $\theta$, which modulates the influence of $U_\text{sim}$ on the score of any game. We sweep over $\theta \in \{0, 0.1, ..., 19.9, 20\}$ (with L1 regularization $\lambda = 0.1$) and compare the best fit aggregate likelihood of the human-created games under the two-stage model to the one-stage model (Figure~\ref{fig:likelihood-score}C).  Incorporating the model-based weighting yields a better total likelihood on the human data, even when accounting for the additional parameter ($\chi$$^2$$=214.6$, $p < 0.0001$; with $\hat{\theta}=5.7$). This suggests that incorporating the model-based simulation is important for explaining the kinds of games people made relative to the broader space of games.

\begin{table*}[ht!]
\centering
\begin{tabular}{p{11cm} c c c}
\toprule
\textbf{Created Game}  & \textbf{$U_\text{sim}$} & \textbf{Percentile} & \textbf{Expressible (Original)} \\
\midrule

\textit{10 $\times$ 10. I call it double trouble! Player one AND player two place two tiles for EVERY turn. The first to connect 5 in a row wins!} & 0.67 & 97.5\% & Yes \\ \midrule 

\textit{12 $\times$ 12. 4 in a row horizontally or vertically wins, or 5 in a row diagonally.} & 0.65 & 96.6\% & Yes \\ \midrule 

\textit{10 $\times$ 10. First player to make either 4 in a row (horizontal, vertical) or a 2x2 square wins.} & 0.61 & 94.1\% & No \\ \midrule 

\textit{6 $\times$ 10. players have to match 4 twice in one game to win the game.} & 0.61 & 94.1\% & Yes \\ \midrule 



\textit{6 $\times$ 6. 2 horizontal or vertical in a row loses} & 0.49 & 66.9\% & Yes \\ \midrule 


\textit{6 $\times$ 6. You must make a 2x2 square block before the other player} & 0.44 & 55.1\% & No \\ \midrule 

\textit{10 $\times$ 10. Alternating turns and number of pieces.  First play, player 1 can play 2 but player 2 can only play 1.  Then P1 can only play 1 but P2 plays 2.  Repeating. P1 plays 2, P2 plays 1. Six in row H/V/D wins.} & 0.42 & 50.8\% & Yes \\ \midrule 

\textit{6 $\times$ 6. You need 5 in a row and you can only win diagonally.  Each player goes twice each time.} & 0.36 & 28.8\% & Yes \\ \midrule 

\textit{10 $\times$ 10. First one to make a plus symbol (3 squares going vertical and 3 squares going horizontal crossing each other) wins} & 0.31 & 18.6\% & No \\ \midrule 

\textit{7 $\times$ 7. Must get 7 in a row to win, only diagonally and second player gets 2 spaces on first turn} & 0.29 & 14.4\% & Yes \\ \midrule 


\textit{12 $\times$ 12. You must get 6 in a row diagonally or vertically.} & 0.16 & 0.8\% & Yes \\ \midrule 

\textit{2 $\times$ 3. each player takes one turn. first to 3 in a row wins} & 0.05 & 0.8\% & Yes \\  

\bottomrule
\end{tabular}
\caption{\textbf{Sample human-created games, of varying funness and ``novelty'' (relative to the DSL used to express the original game context).} Example games varying in $U_\text{sim}$ and whether they are expressible in the restricted subset of the grammar capable of expressing the original $121$ games or beyond. Games are selected to span a range of the kinds of diversity in the games, expressible in \texttt{Ludax}, that were produced by people. ``Percentile'' depicts where the estimated funness of the game ($U_\text{sim}$) falls (as simulated under the Intuitive Gamer model) relative to the distribution of funness over the original set of $121$ games.}
\label{tab:games-w-scores}
\end{table*}

\begin{table*}[t!]
\centering
\begin{tabular}{p{4cm}p{7.5cm}cccc}
\toprule
\textbf{Standard Game} & \textbf{Example Created} & \textbf{$U_\text{sim}$} & \textbf{Percentile} & \textbf{$N_\text{created}$} & \textbf{$N_\text{saw}$} \\
\midrule
10 $\times$ 10, 5 in a row & \textit{10 $\times$ 10. First player to 5 in a row wins. Each player places one square at a time. You can win diagonally, horizontally, or vertically.} \newline \textit{10 $\times$ 10. Five in a row wins. You can move in any direction.} & 0.59 & 94.1\% & 19 & 2 \\ \midrule
10 $\times$ 10, 4 in a row & \textit{10 $\times$ 10  Move one at a time and first to 4 in a row wins}. \newline \textit{10 $\times$ 10. Players alternate turns and the first player with 4 in a row wins.} & 0.63 & 94.9\% & 15 & 2 \\ \midrule
3 $\times$ 3, 3 in a row & \textit{3 $\times$ 3. 3 in a row wins}. \newline \textit{3 $\times$ 3. 3 in a row horizontal, vertical, or diagonal.} & 0.43 & 52.5\% & 10 & 10 \\ \midrule
5 $\times$ 5, 3 in a row & \textit{5 $\times$ 5. The first person to get 3 tiles in a row wins.}. \newline \textit{5 $\times$ 5. 3 in a row wins.} & 0.49 & 67.8\% & 8 & 1 \\ \midrule
5 $\times$ 5, 4 in a row & \textit{5 $\times$ 5. 4 adjacent colors win the game. It can be diagonal, vertical or horizontal.} \newline \textit{5 $\times$ 5. First to 4 wins.} & 0.47 & 64.4\% & 7 & 0 \\ \midrule
6 $\times$ 6, 4 in a row & \textit{6 $\times$ 6  You must be the first to get 4 in a row, any direction}. \newline \textit{6 $\times$ 6. The first player to get 4 in a row wins. Blue goes first and any 4 spots in a row wins, diagonal, horizontal and vertical.} & 0.58 & 88.1\% & 6 & 0 \\ \midrule
8 $\times$ 8, 4 in a row & \textit{8 $\times$ 8  Winner must have 4 pieces in a row, in any direction.} \newline \textit{8 $\times$ 8  four in a row.} & 0.63 & 95.8\% & 5 & 0 \\ \midrule
12 $\times$ 12, 4 in a row & \textit{12 $\times$ 12. Four in a row wins}. \newline \textit{12 $\times$ 12. The goal is to get 4 in a row, including diagonals.} & 0.61 & 94.1\% & 5 & 0 \\ \midrule
5 $\times$ 5, 5 in a row & \textit{5 $\times$ 5. First to 5 in a row wins}. \newline \textit{5 $\times$ 5. You color the cells on the grid. You can hover over a cell. click on a grid to confirm. pieces are either red or blue. the first player to make five in a row wins.} & 0.39 & 41.5\% & 5 & 0 \\
\bottomrule
\end{tabular}
\caption{\textbf{Most frequently human-created games.} The most common games that people created. ``Example Games'' depicts two example participant-written games of the same type as ``Game''. $N_\text{created}$ depicts the number of people who created this game; $N_\text{saw}$ depicts the number of people who saw that game in their context of $11$ games. $U_\text{sim}$ is the Intuitive Gamer model-based prediction for the funness of the game. ``Percentile'' depicts the percentile of $U_\text{sim}$ for that game relative to the distribution of Intuitive Gamer-predicted funness scores for the original 121 set of games.}
\label{tab:game-modes}
\end{table*}




However, not all games are well-captured by accounting for $U_\text{sim}$. We observe a range of games proposed (see Table~\ref{tab:games-w-scores}) with varying base likelihood for having been proposed from their context (e.g., whether or not they are expressible in the restricted DSL used to encode the $121$ games). We discuss alternate explanations for some of the games people proposed in the ``Discussion.'' Additionally, we notice that there are several ``modes'' of games that people seemed to coalesce around (Table~\ref{tab:game-modes}). Many participants created games equivalent to ``\textit{$10 \times 10$ $5$ in a row wins}'' or ``\textit{$10 \times 10$, $4$ in a row wins}'' despite not seeing those specific examples in their prior context. 
Interestingly, these two games are among the most ``fun'' in the base set of 121 games as rated by the Intuitive Gamer model despite being mechanically simple. These modes align with an account of novice game design that involves model-based simulation in the style of the Intuitive Gamer model, though there are of course other unrelated factors that could be at play (e.g. a preference for the number $10$).
It is also notable that people expressed the \textit{same game structure} using different natural language descriptions (see two example participant-written games for each ``game mode'' in Table~\ref{tab:game-modes}). Future work should build off of our early descriptive analyses to better understand whether these repeated examples represent genuine modes in the distribution of human generation and, if so, what mechanisms may be driving people's preference for such games. There are also many interesting questions surrounding the role of communication on human invention \citep{mokyr2010enlightened} and the effects of describing the same underlying games with different natural language.







\section*{Alternate explanations and limitations}

The freeform nature of the games elicited in \cite{collins2026intuitivegamer} coupled with the following limitations constrain the extent to which we can make substantive claims from the present methods about the role of model-based simulation in the people's innovative process at a granular level. While many human-generated games are deemed fun under the Intuitive Gamer, several are not (even excluding trivial or ``silly'' games, see examples in Table~\ref{tab:games-w-scores}). What additional factors might have led people to create the games they did, or how might our current factors not capture the full story?


First, it's possible that people \textit{are} doing model-based simulations as part of generating games but that the details of those simulations don't match our assumptions here. For instance, designers might be simulating under a different ``game world model'' than the one produced by our formalization procedure or might simulate the play of agents that are very different from the Intuitive Gamer model.
Relatedly, novice designers may well be pursuing a model-based notion of ``funness'' but struggle to produce a set of rules that achieves their vision. Indeed, we notice that many participants claim that their designs are balanced between the first and second player but that this is not always borne out by simulated play (see Supplemental Figure~\ref{fig:self-preds-fun-payoff}).
We are actively designing more controlled experiments to better disentangle the roles of representation, model-based simulation, and compute in peoples' innovative decisions in these kinds of games -- and systems of rules and reward more broadly.

Second, novice designers might be using entirely different processes to evaluate a game's engagement even in the absence of explicit simulation.
For instance, a designer might propose a set of rules and conduct a quick check of whether it ``sounds fun'' without actually mentally simulating play in the game, relying entirely on prior experience and linguistic cues in the game description. Consider the following game, submitted by a participant: ``\textit{5 $\times$ 5 board, you color the cells on the grid. You can hover over a cell. click on a grid to confirm. pieces are either red or blue. the first player to make five in a row wins.}'' At first blush, this game appears to implement a variety of potentially engaging mechanics. In practice, however, it reduces to a classic $M-N-K$ game (and one that is very likely to end in a draw under reasonable play). Next steps ought to better explore the role of alternate utility functions that human innovators may bring to bear to their mental evaluative process when deciding what rules to set or change.

Third, our proposal distribution may not fully capture the range of factors driving novice game design. We approximate a game's ``base probability'' using the token log probabilities of a distributional language model conditioned on games people had seen. 
While this choice acts as a useful and general prior, it fails to account for an individual's experiences (both as player and designer), preferences, and tendencies. Future work ought to more deeply explore the unique ways in which a specific individual tackles the problem of invention \citep{kirton1976adaptors}. We also only consider a single LM as a stand-in for how our exposure to games across life affects our production of new games, but this choice -- coupled with our particular method for sampling from the space of possible games -- necessarily introduces a bias in the resulting Bayesian normalization over proposals \citep{bonawitz2010deconfounding}. Our model-based likelihood assessments can also be expanded to assess different ablations of the context participants had seen, e.g., assessing the likelihood of games under all $11$ seen examples (which often included atypical games, like those where a player could play twice on their move, or where they could not win diagonally) with only ``conventional'' games (e.g., Tic-Tac-Toe and Connect Four). 






Finally, we observe that some people created relatively ``silly'' or otherwise non-effortful games. This is somewhat expected, as people were not explicitly incentivized to genuinely create a fun game. We caveat again that our current analyses are an initial empirical look at a large-scale dataset of open-ended human creativity. Future work ought to assess people's innovative choices in more controlled design spaces, and then build back up towards the open-ended design space we explore. The methodological tools we contribute here help unlock such possibilities. 

\section{General discussion}

In life and play, people do not merely follow rules -- they are active participants in the process of revising and producing systems and structures. More than that, people have a sense for what rules and reward structures are reasonable and strong intuitions about which new problems or games may be worth the time to engage with. We build on prior work that explores these intuitions about games both in terms of evaluation and effective play from little experience \citep{lake2017building, dubey2018investigating, tsividis2021human, collins2026intuitivegamer}, and take steps to explore how those intuitions extend beyond problems that are provided to us and into the realm of invention.




While explanatory, our model of game creation does not consider the possibility of \textit{re-sampling} games based on the result of evaluations. A natural next step, then, is to incorporate an explicit ``propose-evaluate-revise'' loop or similar iterative refinement process like evolution \citep{todd2024gavel}. We also aim to more deeply explore the relationships between thinking time and game-related decisions \citep{russek2022time} and the effects on game evaluations of actual (instead of mentally simulated) play. In practice, innovative settings can often involve many stages of strategic decision making~\citep{agrawal2021enabling}, potentially engaging model-based simulation or other kinds of evaluation unrolled over multiple steps.




More broadly, it is worth carefully considering the role that large language models (LMs) play at various stages in our modeling framework.
The study of open-ended game creation is hard, in part as it by definition is open-ended. LMs open up a new way to quantifiably study creativity over a wide range of natural language game descriptions, fundamentally allowing us to perform these analyses in a way that was previously impossible.
Concretely, our model-based analyses rest on the ability to have a model in which to simulate play; language-to-code models allow us to synthesize such ``game world models'' at scale~\citep{wong2025modeling} for open-ended modeling of participant-generated text. However, the interpretability of LMs as cognitive models (or parts of cognitive models) remains an open and active area of discussion~\citep{sucholutsky2025using, mccoy2024embers, yildirim2024task, wong2025modeling, rmus2025towards,binz2025should}, and we look forward to deeper engagement with alternate modeling formalisms as we scale the study of open-ended invention, powered in part -- as we show here -- via model-based simulation of play in the invention (here, game). 


Our work is only a first step in a model-based analysis of the remarkable capacity that people have for invention and innovation.
The computational tools and techniques we deploy here are powerful but incomplete: we are excited about the possibility of strengthening and sharpening them as part of more empirical studies designed specifically to tease apart the role and relationships between the proposal and evaluation processes in human invention. Much remains to be understood about how people shape the systems of rules with which they interact and invent entirely new ones.

\section*{Acknowledgements}

We thank Lance Ying, Prafull Sharma, Alex Lew, Scott Stern, and Mauricio Barba for helpful conversations that informed this work. This work was supported by AFOSR (YIP FA9550-23-1-0127), the ONR Science of AI program (N00014-23-1-2355), a Schmidt AI2050 Fellowship to JBT, and the Siegel Family Quest for Intelligence at MIT. KMC acknowledges support from the Cambridge Trust and King's College Cambridge. LCW acknowledges support from a Stanford HAI Fellowship. AW  acknowledges  support  from  a  Turing  AI  Fellowship  under grant  EP/V025279/1, The Alan Turing Institute, and the Leverhulme Trust via CFI. This work is supported (in part) by ELSA - European Lighthouse on Secure and Safe AI funded by the European Union under grant agreement No. 101070617. Views and opinions expressed are however those of the author(s) only and do not necessarily reflect those of the European Union or European Commission. 

\bibliographystyle{apacite}

\setlength{\bibleftmargin}{.125in}
\setlength{\bibindent}{-\bibleftmargin}

\bibliography{references, refs_games}

\newpage

\appendix


\section*{Additional Details on Human Game Invention Process and Created Games} 

\subsection*{Creation Time and Scratchpad Usage}
\begin{figure}[h!]
    \centering
    \includegraphics[width=0.8\linewidth]{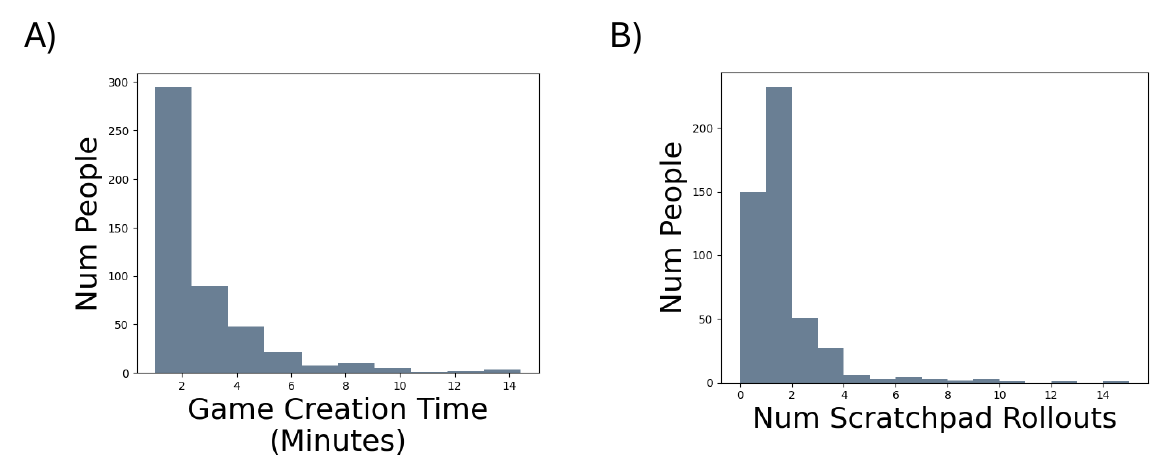}
    \caption{\textbf{Descriptive statistics on people's game creation process.} (A) Time participants spent to create their game (minutes). (B) Number of ``rollouts'' any one participant made with the interactive scratchpad provided. Number of rollouts indicates the number of ``fresh'' explicit traces a participant made on the game (based on the number of times they pressed the ``clear'' button on the board).}
    \label{fig:game-time-creation-human}
\end{figure}

Figure~\ref{fig:game-time-creation-human} includes additional details on people's game creation process, including their time to create the game and the amount of interaction they had with the scratchpad. Participants were required to spend at minimum one minute creating their game; nothing required a participant to spend \textit{more} than one minute. Approximately 50\% of participants spent two or more minutes creating their game. During this time when participants were thinking about their game, they were permitted to engage with an example gameboard ``scratchpad'' while creating their game. The scratchpad was fixed at a $13 \times 13$ board. The scratchpad could be used to simulate play by making moves for both players and manually resetting the board. Participants also were not required to use the optional scratchpad provided when creating there game; however, many did (some even conducting multiple rounds of click-and-clear). This may indicate external simulation of play; we are actively exploring the role of scratchpad in relation to people's tendency to simulate play. Additional methodological details can be found in \cite{collins2026intuitivegamer} and \cite{collins2026intuitivegamer}.

\subsection*{Games Expressible in the Restricted Grammar}

\begin{figure}[t!]
    \centering
    \includegraphics[width=0.8\linewidth]{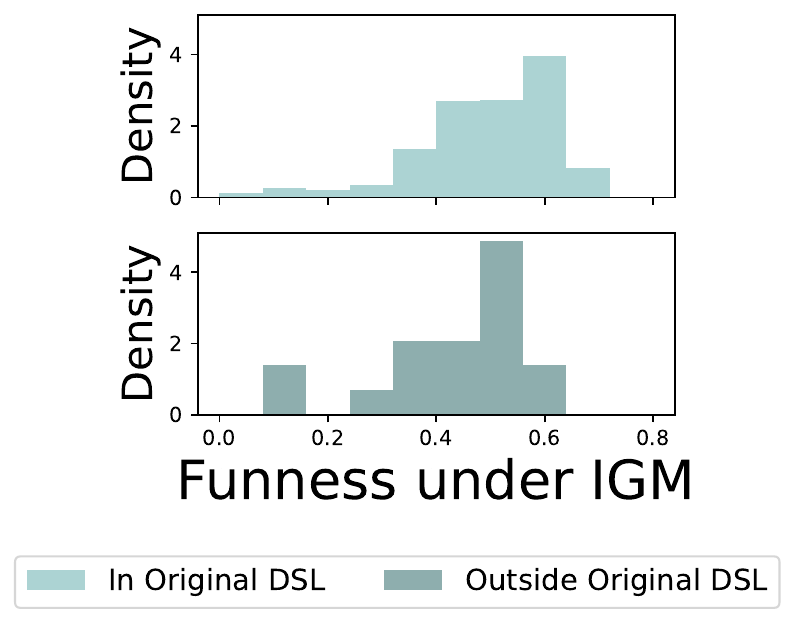}
    \caption{\textbf{Predicted model funness, by approximate novelty.} Intuitive Gamer-model predicted funness scores for human-created games that are expressible in the original restricted grammar used to express the context of $121$ games (top) compared to the scores for the subset of \texttt{Ludax}-expressible games people created that incorporate DSL features that go outside of those that can express the original $121$ set (bottom).}
    \label{fig:novel-fun-dsl}
\end{figure}

Figure~\ref{fig:novel-fun-dsl} shows the predicted funness scores, under the Intuitive Gamer model, for games that participants created which are expressible using only the restricted grammar subset of \texttt{Ludax} used to create the original $121$ games compared to the games that are expressible in \texttt{Ludax} but use DSL component(s) outside of the restricted grammar. The latter games can be considered more ``novel'' relative to the games people saw. The games outside of the restricted set are not necessarily more fun than games that combine or otherwise engage only with DSL components expressible in the original set of games. It is possible that people who thought more ``out of the box'' went too far; or, it may be that the Intuitive Gamer funness model (which has been well-validated on the original $121$ games) does not as well-generalize to games that are more distant from the base set of games. Or, it may be that people who thought further out of the box did not simulate at all, or less. Future work can better investigate such possible explanations and the role going beyond context in generation with structured evaluation.

\subsection*{Games Inexpressible in \texttt{Ludax}} 

Some participant-generated games could not be expressed in the version of \texttt{Ludax} used for these experiments. In some cases, this was because the game descriptions themselves were underspecified (e.g. they did not properly define the game's dynamics or end conditions). In other cases, however, the games were perfectly understandable in natural language and could only not be expressed programmatically due to technical limitations. The \texttt{Ludax} description language is defined over two-player, deterministic, perfect-information board games -- games which violate one or more of these constraints cannot be expressed in the DSL at all. This is an inherent limitation of this work, and we aim to increase the range of games captured by our simulation step in future efforts. In \autoref{tab:inexpressible-games} we present a sample of participant games that are inexpressible in \texttt{Ludax}.

\begin{table*}[ht!]
\centering
\begin{tabular}{p{9cm} p{7.5cm}}
\toprule
\textbf{Game Description}  & \textbf{Reason Not Expressible} \\
\midrule
\textit{Draw a object using the colored squares but only using a specified amount of squares} & Game is underspecified. \\
\midrule
\textit{You have to get 4 in a row and yell Floofy before you win or else it doesn't count} & Game refers to something outside the simulated board. \\
\midrule
\textit{Both players place boxes as fast as they can and the first to have 5 same color boxes wins.} & Game is not turn-based. \\
\midrule
\textit{The first person to get three in a row using the center squares only wins. The other person (opponent) doesn't know this.} & Game relies on hidden information. \\
\midrule
\textit{2 players flip a coin to decide who goes first. First Player always wins.} & Game is not deterministic. \\
\midrule
\textit{You can win by getting 5 in a row in a vertical or horizontal pattern, or by getting 4 in a row diagonally.  Player 1's first move includes placing two pieces, but the second player has one opportunity during the game to remove one of player 1's pieces.} & DSL does not support ``once per game'' actions. \\

\bottomrule
\end{tabular}
\caption{\textbf{Samples of human-created games yet not expressible in the current version of \texttt{Ludax}.} The \texttt{Ludax} description language does not cover all possible games or even all possible board games -- it is restricted to a subset of two-player, deterministic, perfect-information games. Some participant-generated games are plausible and well-defined, but cannot be expressed programmatically because of these technical limitations.}
\label{tab:inexpressible-games}
\end{table*}


\subsection*{Predicted Payoff under Intuitive Gamer}

\begin{figure}[t!]
    \centering
    \includegraphics[width=0.85\linewidth]{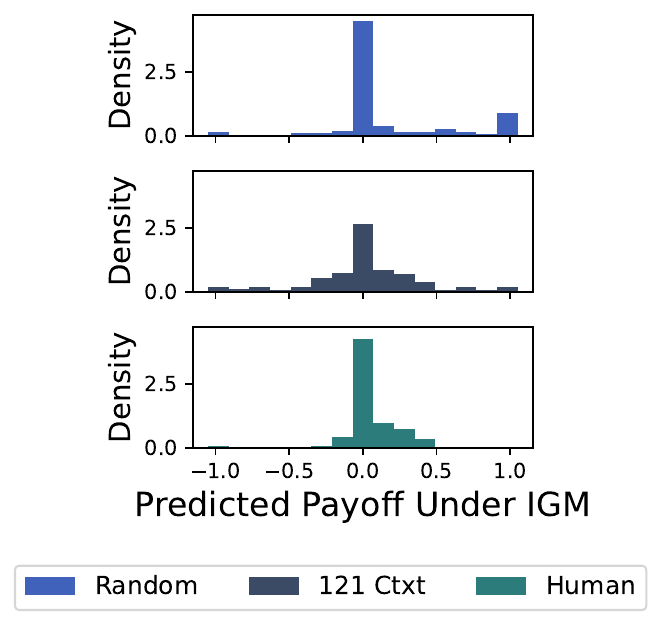}
    \caption{\textbf{Predicted payoff under simulated novice play}. Expected payoff computed under Intuitive Gamer simulations, for games produced from three sources: randomly sampled from a DSL over the space of games (top); the original $121$ games participants may have seen (middle); and the human-created games that are expressible in \texttt{Ludax} (bottom). Bootstrapped per-game averaged payoff, computed using $k=6$ simulated games bootstrapped averaged under $20$ simulated participants, to mimic the original evaluations of \cite{collins2026intuitivegamer}.}
    \label{fig:model-pred-payoff}
\end{figure}

The expected payoff of simulated games in the games people created (over the \texttt{Ludax} representations) is generally near zero (Figure~\ref{fig:model-pred-payoff}). That is, people tended to create fair games that are predicted to be fair (when played by novices). This matches participants' own judgments that their games were likely to be fair as we outline below. Arbitrarily sampling from the space of games tends to also produce games with an expected payoff of zero. It is noteable that even though both human-generated and randomly-generated games tend to have similar expected payoffs (``fairnesses''), the games people make exhibit some structure that causes them to be evaluated as more fun under the Intuitive Gamer model. 


\subsection*{People's Self-Evaluations}

We focused the bulk of our main analyses on model-based evaluations of the games people created. However, after people created a game, they either rated the expected funness or expected outcomes of the game (matched to the questions they were asked about for their evaluations of the first set of $11$ games). We next conduct exploratory analyses into each group of participants' self-ratings based per question type (funness and outcomes).

\begin{figure}[t!]
    \centering
    \includegraphics[width=1.0\linewidth]{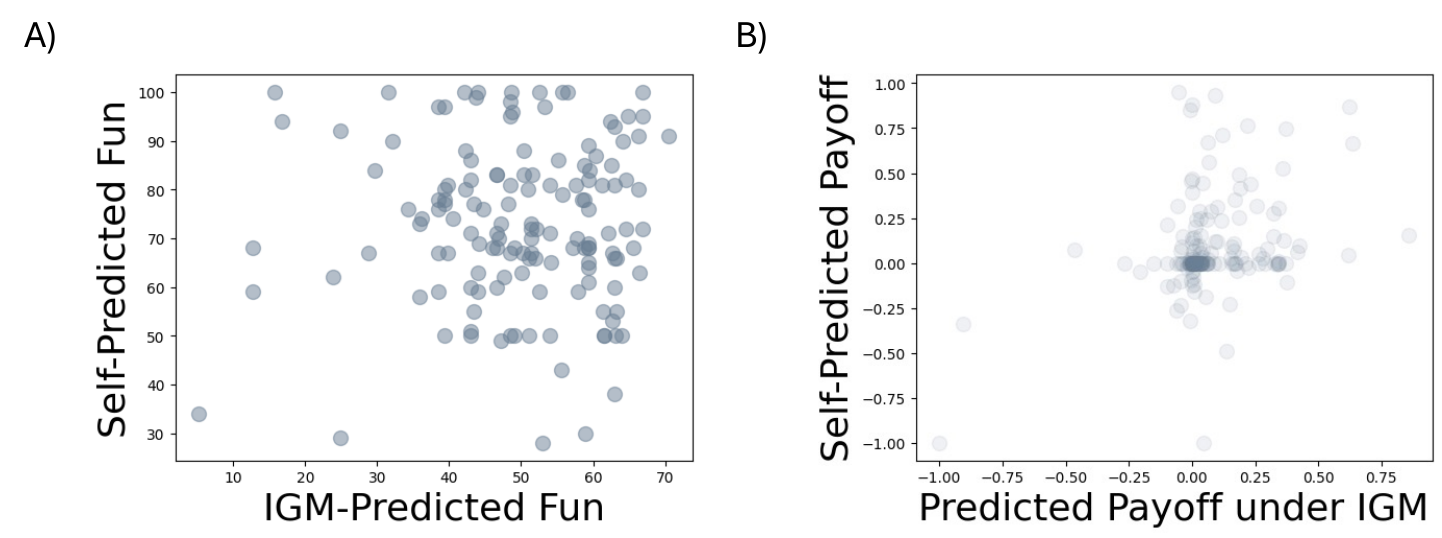}
    \caption{\textbf{People tended to think their games would be fun and fair.} Participants rated their games as generally fun (A) and fair (i.e., payoff zero; B). Participants' payoff predictions were highly aligned to the expected payoff under Intuitive Gamer simulations (i.e., near zero). Participants' funness judgments were more varied relative to Intuitive Gamer predicted funness.}
    \label{fig:self-preds-fun-payoff}
\end{figure}

\paragraph{Funness predictions}
People generally thought their games would be fun, which was not well-correlated with the expected funness under the Intuitive Gamer model ($R^2$ near zero; see Figure~\ref{fig:self-preds-fun-payoff}A). It is expected that most people would have an inflated sense of the funness of their game, as people were asked to make a fun game. Yet, it is interesting that several people did not rate their likely game as being extremely fun. The tendency for some people to think their game may not actually be that fun may also illustrate the challenge of generating a fun game versus evaluating an existing game. 

\paragraph{Outcome predictions}
People who created their game tended to rate their game as likely to be fair. This is interesting as people were only asked to make a fun game -- not necessarily a fair game. Yet, many people thought their game would be fair (i.e., payoff zero). People's predictions were generally aligned with the expected payoff under the Intuitive Gamer simulated outcomes that the games should yield payoff zero (see Figure~\ref{fig:self-preds-fun-payoff}B).  

\begin{figure*}[t!]
    \centering
    \includegraphics[width=1.0\linewidth]{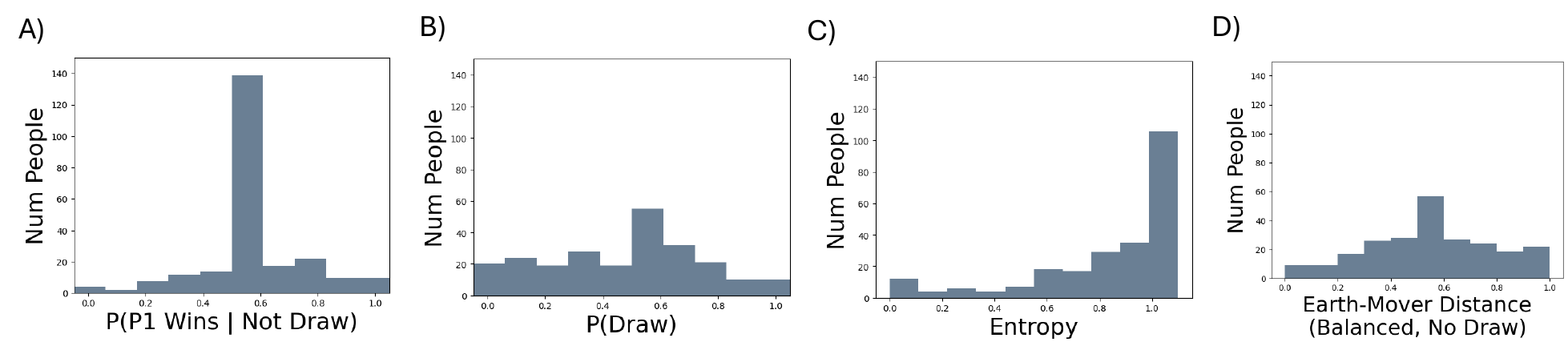}
    \caption{\textbf{People's predictions for the likely outcomes of the games that they created.} After participants created their game, approximately half of the participants were asked to predict the expected outcomes of play in their game (by two reasonable players). Participants were asked to predict how likely the first player would win given the game does not end in a draw (A) and how likely the game is to end in a draw (B). From these measures, the expected payoff can be computed, as well as other measures over the space of outcomes, e.g., entropy over the likelihood that the game ends in a loss/draw/win (C), or the Earth Mover's Distance (``balance'') between the likelihood that the game has equal odds of Player 1 and 2 each winning (and definitely not ending in a draw; panel D). If all outcomes are equally likely, then the entropy is zero (C). Higher Early Mover Distance indicates that the likely distribution over game outcomes (as predicted by the player) is farther from equal odds of Player 1 and 2 winning and no draw (D).}
    \label{fig:self-preds-outcomes}
\end{figure*}

Notably, a game can attain an expected payoff of zero via several paths: all games may end in a draw; all outcomes may be equally likely (P1 wins, draw, or P2 wins); or, P1 or P2 may win in equal combinations (but no draw). Our Intuitive Gamer funness model focuses on the latter: how balanced a game is wherein the game is not likely to end in draw. This encodes the notion from ~\cite{collins2026intuitivegamer} that people favor games that have low draw rates. However, creating a game with perfect balance between P1 and P2's odds of winning without draws may be challenging to create. People tend to believe that the intended-fun game they created has equal odds for P1 and P2 (Figure~\ref{fig:self-preds-outcomes}A), but with still some moderate probability of a draw in any playthrough (Figure~\ref{fig:self-preds-outcomes}B). This leads to games that have generally high entropy (Figure~\ref{fig:self-preds-fun-payoff}C) and perhaps are surprising as a result; yet, score relatively poorly on the expected ``balance'' component (which penalizes draws; Figure~\ref{fig:self-preds-fun-payoff}D). Why is it that people tend to rate games as likely fun if they have low draw rates, yet create games that they propose (and recognize that they have proposed) with expected non-neglible draw rates? Is it from the challenge of creating games that have low draw rates -- wherein participants at minimum try to create a fair game for it to be fun, thereby settling with draws? Future work can better investigate ease of generating fun games versus evaluating given games' likely funness. 

\subsection*{People rating the games people made and creating new games}


People rated the games they created, but how fun do others find the games they made? And participants in the first study created games after evaluating a provided set of games: what kinds of games do people create in a second round of innovation,  conditioned on samples from this first round? We take initial steps to explore these questions with a follow-up study with a new group of human participants.

\paragraph{Human evaluation of the games people made} 

The game creators' goal was to create a game that would be fun. To assess the extent to which the games people created achieved this goal, we recruited $227$ participants\footnote{We remove $14$ participants who only used the endpoint or exact middle of the slider on over $70\%$ of trials.} to \textit{rate} whether the games created by people are fun using the same experimental set-up as in ~\cite{collins2026intuitivegamer}. That is, this new group of participants were presented with the game description (board size and game rules) as well as an interactive game board over which they could optionally interact before determining their rating. Participants were allowed to click on the board (and change move colors to reflect player changes) as well as undo and reset clicks. Participants judge $10$ games: Tic-Tac-Toe (to calibrate ratings against our original set of ratings) as well as nine games from the human-created games studied in the main text. We randomly sampled the nine games from buckets based on the time the game creator took to create the game (we sample three games from the games that took less than $1.5$ minutes, between $1.5$--$3$ minutes, and over $3$ minutes, respectively). We collect ratings for a total of $108$ of the human-created games. After rating all of the games, participants are again asked to create a new game that others would find fun. They then rated their created game on a slider indicating how fun they find the given game relative to this class of grid-based games (``least fun of this class of grid-based games'' up to ``most fun of this class of grid-based games''). This produced a new set of $213$ created games, which we give a preliminary analysis of in the final section.  

\begin{figure}
    \centering
    \includegraphics[width=0.7\linewidth]{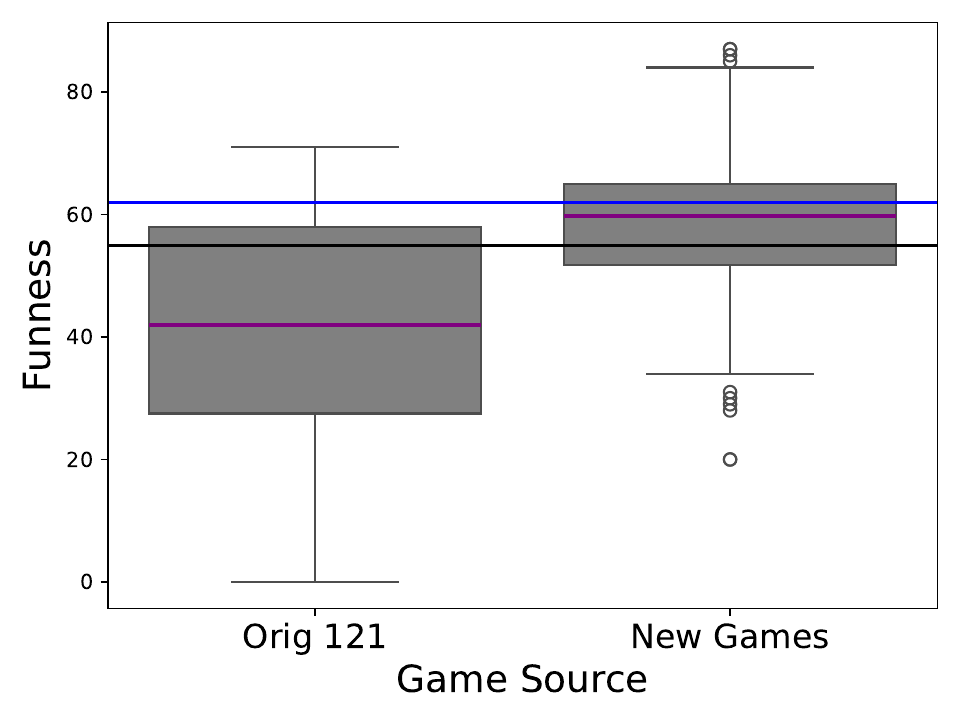}
    \caption{\textbf{Human-created game funness ratings compared to original game ratings.} Funness scores for the games in the original batch of $121$ games versus a new set of raters on a subset of $N=108$ of the games that people created. Each group also rated Tic-Tac-Toe. We compute the median funness score for each game; blue indicates the median funness in our new rating set for Tic-Tac-Toe compared to black in the original ratings in \cite{collins2026intuitivegamer}. } 
    \label{fig:funness-old-v-new}
\end{figure}


\paragraph{People tend to rate other people's games as likely fun, relative to the original batch of games.}


The games that people create also tend to be rated as \textit{more fun} than the original batch of 121 games that had provided the context for that the game innovators saw when they created their game. The median funness rating in our new ratings is $59.8 \pm 2.5$ (out of $100$) compared to $42 \pm 3.5$ in the original batch of 121 games. We note some caution in comparing across distributions, as the first game set had many more deliberately unfun games; we therefore expect that it is possible the new games would be even \textit{more} fun (as we have a lower calibration window). We see that Tic-Tac-Toe received a slightly higher absolute funness score ($62 \pm 3.9$ standard error) in the new batch compared to $55 \pm 3.7$ in the original batch (though is relatively closer to the median score), somewhat suggestive of this point. The split-halve $R^2$ between participants is only moderate, suggestive of general variability across participants' ratings (mean $0.42$, 95\% CI: [$0.33, 0.54$]). We suspect that some of the discrepancy in funness ratings is driven by participants' responsiveness to the linguistic elements of the game description -- they may be more likely to rate a game described in exciting language as fun than a programmatically equivalent goal described in comparatively duller standardized language.


\paragraph{Humans create games that are even more novel relative to the original $121$ set after a second iteration of creation.}

The participants who rated the human-created games also then created their own new game, producing one additional round of ``iteration'' in the invention process. We again see that context influences the games people create: 67\% of the newly created games are expressible in the restricted grammar compared to over 80\% of the original batch of new games (as translated by the LM). However, crucially: as the context of these games are now more novel, we see that an even smaller percent of games are expressible in the original restricted grammar of $121$ games. In exploratory analyses, we again see the effect of creation time on novelty, but now moving even further (across the population) from the original set of games of $121$ games: 77\% of games created by people in $< 1.5$ minutes ($N=39$) are expressible in the original restricted grammar compared to 51\% of people who took between $1.5-3$ minutes ($N=89$) and only $41$\% expressible for games where the creator took $> 3$ minutes ($N=85$).

\begin{table*}[ht!]
\centering
\begin{tabular}{p{11cm} c c c}
\toprule
\textbf{Created Game}  & \textbf{$U_\text{sim}$} & \textbf{Percentile} & \textbf{Expressible (Original)} \\
\midrule
\textit{8$\times$8. If you make 4 in a row (horizontal or vertical only) you lose. Try to win by forcing your opponent to make 4 in a row.} & 0.71 & 100.0\% & Yes \\ \midrule
\textit{9$\times$9.  First one to get 3 in a row diagonally losses otherwise the game is a draw.} & 0.69 & 98.3\% & Yes \\ \midrule
\textit{8$\times$8. First player with 3 pieces connected to form a partial square are eliminated} & 0.63 & 95.8\% & No \\ \midrule
\textit{10$\times$10. First to form an L-shape wins} & 0.62 & 94.9\% & No \\ \midrule
\textit{10$\times$10. The first person to link five consecutive lines horizontally, vertically, or diagonally—wins. Special Rules: Players have just 5 seconds to move. Two blocker pieces are given to each participant. Also, turn is granted by a move that creates two rows of four.} & 0.59 & 94.1\% & Yes\footnote{Interestingly, this game was translated in the formalization step simply to $10\times10$, $5$ in a row wins, highlighting the need to expand the quality of the formalization step to cover the kind of innovative games participants come up with.} \\ \midrule
\textit{10$\times$10. first player to get a square of 4 colors together wins. each player may only place one square at a time} & 0.53 & 77.1\% & No \\ \midrule
\textit{10$\times$10. The player who makes a 2x2 square first wins} & 0.53 & 77.1\% & No \\ \midrule
\textit{10$\times$10. First to connect 4 only diagonally. If you accidentally make a connect 4 horizontally, you lose.} & 0.53 & 77.1\% & Yes \\ \midrule
\textit{8$\times$8. Players must create a plus sign -- a 3 vertical and 3 horizontal arrangement with a shared square in the center.  Players get 2 pieces per turn.} & 0.50 & 71.2\% & No \\ \midrule
\textit{6$\times$6. 1) Any player that lands 6 cells in a row, column or diagonally wins..2) Any player that has 3 cells lined up in a row gets an extra turn.} & 0.42 & 50.8\% & Yes \\ \midrule
\textit{12$\times$12. First to make a 5 diagonal row ``x'' wins} & 0.24 & 5.9\% & No \\ \midrule
\textit{12$\times$12. You must get two sets of 6 in a row to win the game.} & 0.17 & 0.8\% & Yes \\
\bottomrule
\end{tabular}
\caption{\textbf{Sample second-round created games.} Selection of the games that people created in the second-round of generation (when a new group of participants were tasked with creating a fun game, after having rated the funness of other people's created games). Games are selected to span a range of the kinds of games and expected funness. ``Percentile'' and ``Expressible'' are relative to the original set of $121$ games and give a measure of the funness relative to the initial set (``Percentile'') and whether the proposed game is representable in the restricted subset of the \texttt{Ludax} grammar that can fully define the $121$ games. A percentile of 100\% means that the game exceeded the highest funness score in any game in the $121$ initial set.}
\label{tab:games-w-scores}
\end{table*}

We conduct an initial exploration into whether these games too may be fun to play. We again concretize the natural language games into \texttt{Ludax}. Approximately $64\%$ of games translate into valid \texttt{Ludax}, which makes sense as the games are generally less expressible in the respective DSL (as noted above), potentially as they move even further beyond the games that they were conditioned on. We score the resulting translations under $P_\text{propose}$ (see details below) and simulate play under the Intuitive Gamer model. The resulting games are generally somewhat less likely under the base distribution than the first round of human-created games. This is expected as the second round of human-created games have been conditioned on other (human-created) games that have moved beyond their context. Interestingly, the games are distributionally highly-similar under expected funness. Both distributions of human-created games similarly move beyond the expected funness (in general) of the original $121$ games and games that would otherwise be randomly sampled from the space of possible games. However, the human-created games in the second round are not notably more fun (according to $U_\text{sim}$). 

We again caveat these results as preliminary, as they operate over \texttt{Ludax} representations that have been freely translated from natural language and therefore warrant further verification that they are aligned to the raw participant-proposed natural language game description. Nonetheless, we are excited by future work, using the language-computable paradigm we develop here for studying iterative innovation in systems of rules and rewards.


\begin{figure}
    \centering
    \includegraphics[width=1.0\linewidth]{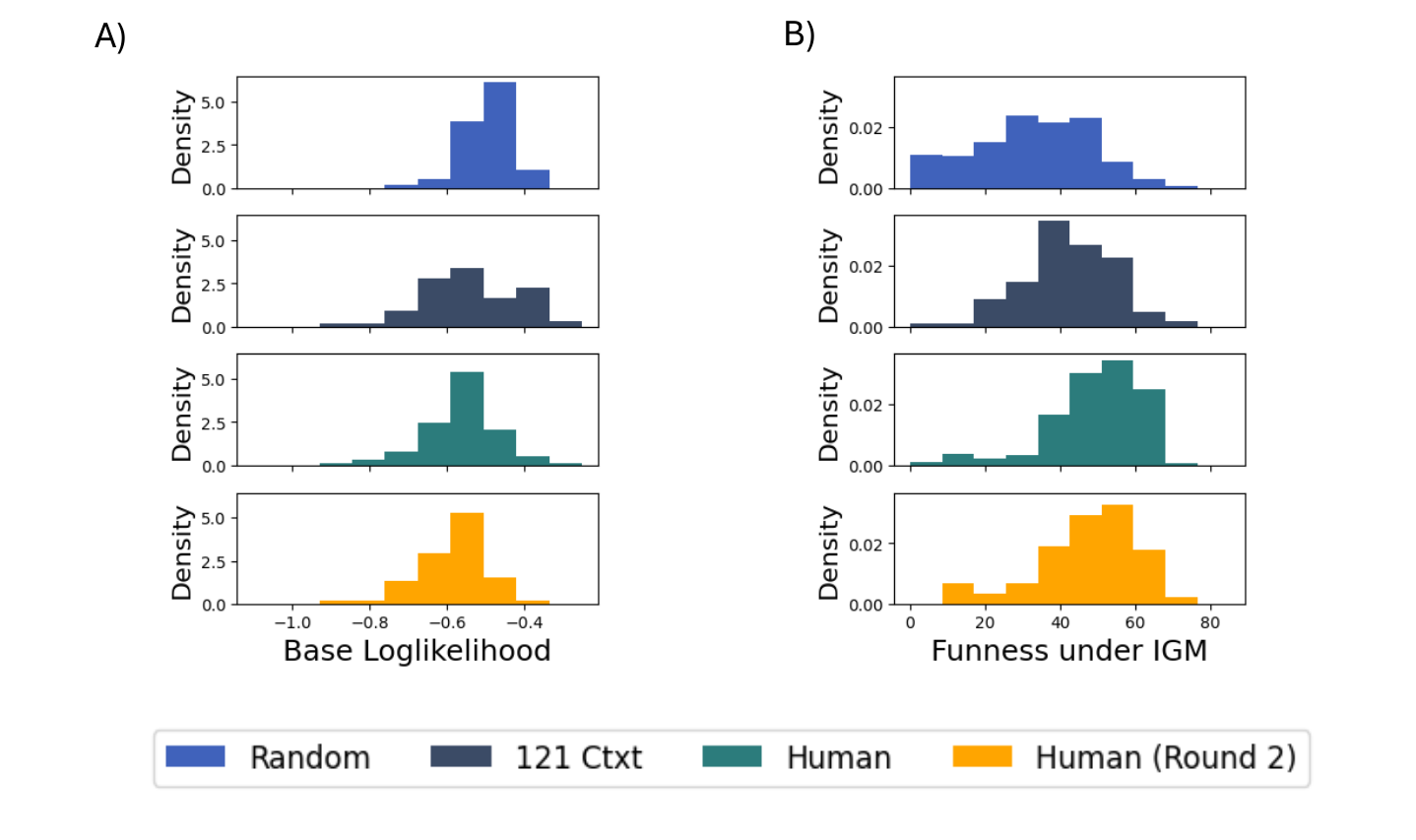}
    \vspace{-0.5cm}
    \caption{\textbf{Likelihood of second round of human-generated games under game creation processes, compared to the first round and baseline comparator games.} As in Figure~\ref{fig:likelihood-score}, (A) depicts the likelihood of each game's program representation (encoded in \texttt{Ludax} is scored under $P_\text{propose}$ and reported as the average token log probability. (B) depicts the average Intuitive Gamer-predicted funness score based on simulating novice play in each game. The games a second round of participants created (bottom) are generally predicted to be similarly fun to the games people created in the first round (third row) and are generally less likely under $P_\text{propose}$ where $P_\text{propose}$ is conditioned on samples from the first $121$ context.} 
    \label{fig:likelihood-decomp-round2}
\end{figure}

\section*{Additional Details on \texttt{Ludax}}

\subsection*{Restricted Grammar}

To create the ``restricted grammar'' that spans the original set of 121 games, we include (potentially asymmetric) line completion games of arbitrary orientation (e.g. ``make a horizontal line of 4''), arbitrary player move ordering (e.g. ``Player 1 opens twice, then players alternate making moves''), and win-inversion (e.g. ``making 3 in a row loses''). We note that the resulting DSL is still compositional and \textit{can} be used to express games not in the original set. Compared to the full \texttt{Ludax} grammar, the restricted DSL notably excludes score-based games, shape-completion games, and a variety of low-level functions and predicates (e.g. detecting the corners of the board).

\subsection*{Sampling from the DSL}


To compute a likelihood of any game in the presence-only MaxEnt model we consider here, we need to have a broader space of games to normalize against. We sample from a DSL defined over games to approximate this normalization. Specifically, each game is obtained by first randomly sampling a board width and height from $\{1, 2, ... , 12\}$. We then sample one or more ``modifications'' derived from the semantic game categories outlined by \cite{collins2026intuitivegamer}. Specifically, we independently sample whether the players have the same or different target line length $K$ (and then sample a value for $K$ based on the board size), whether one or both players are restricted in the directions of lines they can use to win (e.g. diagonal, orthogonal), whether one player opens the game by moving twice in a row, and whether the game is played for a loss instead of a win. In each case, we sample the ``deviation'' (i.e., the result which would lead to a different mechanic than is present in Tic-Tac-Toe) with probability $1/10$ to roughly align with the presence of those mechanics in the original dataset. We rejection sample to ensure $1000$ unique games.

\subsection*{Compute-Bounded Gameplay Model}

Following the general procedure of \cite{collins2026intuitivegamer}, we obtain estimates of game balance and length using $200$ playouts of the Intuitive Gamer model against itself with a move sampling temperature of $1$. Balance is computed as the Earth Mover's Distance over the outcome distribution compared to a distribution where Player 1 and Player 2 have equal chances of winning (and the game does not end in a draw). To obtain the estimate of ``challenge,'' we use $100$ playouts of the Intuitive Gamer model against a random agent with a sampling temperature of $0.2$. Features of simulated play are used to compute the final funness score (Figure~\ref{fig:method-overview}). All playouts are done on the \texttt{Ludax} translated game, if expressible (games that are not expressible in \texttt{Ludax} are excluded from our model-based analyses). In this preliminary work, we use only the averaged coefficients (fit to participant funness ratings for the $121$ games) over the readouts for balance, challengingess, and game length from \cite{collins2026intuitivegamer}. Future work could instead estimate a distribution over possible $U_\text{sim}$ using, for instance, sampling over a distribution of coefficients for each readout. 





\section*{Additional Details on $P_\text{propose}$}

Our hypothesis for how people innovate new systems of rules and reward is that their choices for what to change, modify, or create wholesale are partly based on systems (in our case, games) they have seen before as well as model-based simulation. This is a big question, and we take initial steps to investigate it computationally -- on a large open-ended dataset of human-created games. As such, we make several initial simplifications. 

For instance, $P_\text{propose}$ is meant to capture the a priori probability of sampling a particular game, given the game context, which we take a first approximation of $P_\text{propose}$ by scoring the base loglikelihood of participants' generated games conditioned on the games they had seen. However, as noted, participants have substantially varied natural language surface form of the games they wrote. To standardize the form, we take the average token log probability of the \texttt{Ludax}-representations of the game, given the context games also expressed in \texttt{Ludax} form. This means that we only score the human-created games that are translatable in \texttt{Ludax}. We prompt the model with an abbreviated version of the human participant experiment instructions (see below). 

To approximate the base likelihood of the normalization games (the original $121$ and games sampled arbitrarily from the grammar), we randomly sample from the base set of $121$ games -- sampling $10$ games plus Tic-Tac-Toe to match the context provided for human-generated games. For the original $121$ games, we ensure that the same generated game is not re-included in the prompt as an example translation. We apply the same context sampling for the second round of human-generated games. As the second round human-generated games were conditioned on the first round of games participants created, ideally, we would condition on the translations of those games. However, some games did not have corresponding \texttt{Ludax} translations (as noted above); for direct comparison, we condition only then on translations from the original $121$. 

We use LLaMA 3.1 8B with temperature $0.5$ to score the token level logprobabilities of the game translation. The aggregate score for each game is the average token log probability to avoid length effects. 

Future work can explore alternate context conditioning methods (e.g., including all of participants' past ratings of the games) as well as comparing against variants conditioned directly on natural language (rather than \texttt{Ludax}). Sensitivity of the scores to other LMs and temperature are also important next steps to robustly assess the role of context on the expected generation probability of a game.


\subsection*{Prompt}

\begin{promptbox}
Propose a game that can be played on a grid, similar to games like Connect 4, Gomoku (5-in-a-row), or Tic-Tac-Toe.

Specifically, create your own game that you would find fun to play!

We ask that you provide: 
- The board size
- Rules to win and any other special conditions of your game.
    
The games should be reported in a symbolic programmatic form. 

<<EXAMPLE_LUDAX_GAMES_FROM_CTXT>>

Now, create your game. Use the same format.
\end{promptbox}

\section*{Additional Details on LM-Based Formalization}

\begin{figure*}[t!]
    \centering
    \includegraphics[width=\linewidth]{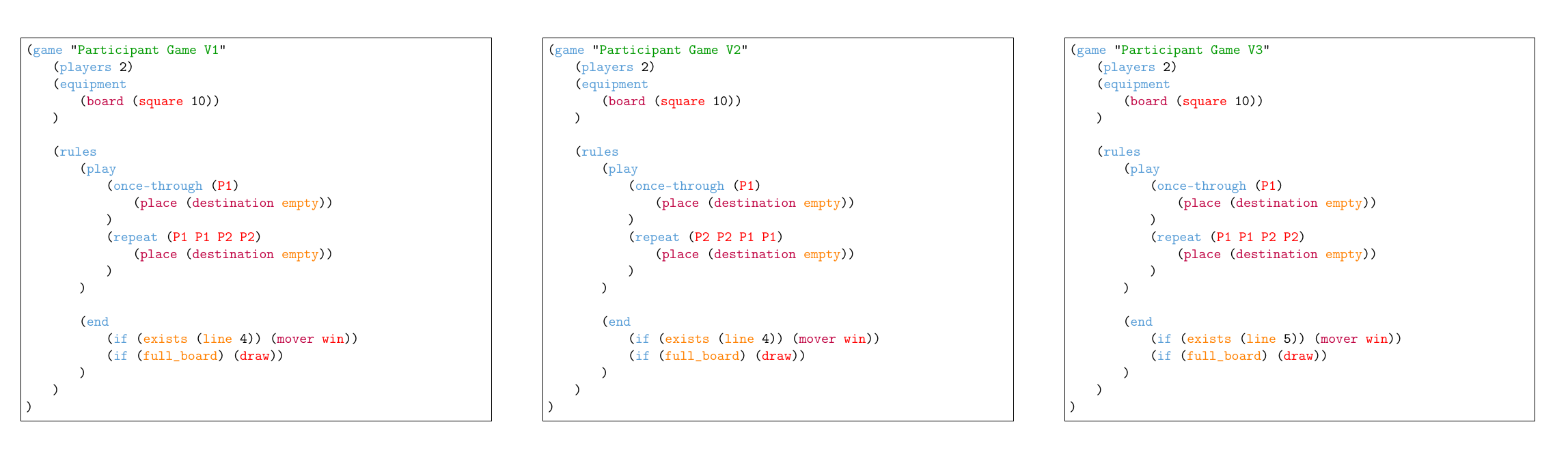}
    \caption{\textbf{Ambiguity and formalization.} Three potential formalizations of the ambiguous game ``10 $\times$ 10 board, Player 1 goes once then after that each player goes twice in a row,'' generated by our formalization procedure. The target line length is unspecified -- two translations (left and center) set $k$ to 4 and one (right) sets it to 5. In addition, two translations (left and right) interpret the ``twice in a row'' phase as beginning with Player 1, causing them to actually open with three moves in a row, while the remaining translation (center) has Player 2 move twice after Player 1 opens.}  
    \label{fig:ambiguous-games} 
\end{figure*}

Participants provided open-ended arbitrary game descriptions in natural language. To assess the potential role of model-based simulation, we needed to translate the games to a symbolic representation. We use LLaMA 3.3 70B, prompted with gold-label translations of $48$ (roughly $10\%$) of the human-created games to attempt to concretize the open-ended natural language game descriptions participants wrote. In addition, the formalization model received an annotated version of the DSL grammar in the system prompt and was instructed indicate when a game cannot be represented in the grammar. To account for uncertainty in the translation process, we sample $k=3$ translations from the LM at a temperate of $0.2$ and average the results over all syntactically-valid translations. Of the games that are annotated by one of the authors to be codeable in \texttt{Ludax}, approximately 90\% (294 of 326) compile; 32 do not compile. Of the games coded to be not expressible in \texttt{Ludax}, none of the translated games compile -- indicative of high-precision in not hallucinating in the \texttt{Ludax} formalization step.  We do notice several cases ($N=29$) of ambiguous games. Figure~\ref{fig:ambiguous-games} presents a set of example translations which collectively span multiple interpretations; however, we may expect that the ``default'' $K$-in-a-row is actually $K=3$ (Tic-Tac-Toe) for this particular game set. Future work should better explore how to model the kinds of game world models people may have imagined when ambiguous in their expression, and ensure the formalization procedure permits similar uncertainty over translations. 

\subsection*{System Prompt}

\begin{promptbox}
You are an expert programmer.
You are tasked with converting a game, expressed in natural language, to a formal representation of the game.

Here is the DSL of the representation:

<<DETAILS_ON_DSL>>
\end{promptbox}

\subsection*{User Prompt}

\begin{promptbox}
You are an expert programmer.
You are tasked with converting a game, expressed in natural language, to a formal representation of the game.

Provide your translation as the last line of your response, in the following form
Translation: <your-translation>

If the game is not translatable in the DSL, instead output CANNOT TRANSLATE. 

Note that games which only specify a condition for losing (rather than winning) can be translated by using the lose motif.

<<EXAMPLE_TRANSLATED_GAMES>>

Now, translate the following:

<<NL_GAME>>"

\end{promptbox}

\end{document}